\documentclass[pre,superscriptaddress,showpacs,preprint]{revtex4}
\usepackage{amsmath,graphicx,amssymb,amsfonts}
\begin{document}

\title{Proportional Paths, Barodesy, and\\ Granular Solid Hydrodynamics}
\author{Yimin Jiang}
\affiliation{Central South University, Changsha, China 410083}
\author{Mario Liu}
\affiliation{Theoretische Physik, Universit\"{a}t T\"{u}bingen, 72076 T\"{u}bingen,
Germany}
\date{December 16, 2012}

\begin{abstract}
{\em Propotional paths} as summed up by the {\em Goldscheider Rule} ({\sc gr}) -- stating that given a  constant strain rate, the evolution of the stress maintains the ratios of its components -- is a characteristics of elasto-plastic motion in granular media. {\em Barodesy}, a constitutive relation proposed recently by Kolymbas, is a model that, with {\sc gr} as input,  successfully accounts for data from soil mechanical experiments. {\em Granular solid hydrody\-namics} ({\sc gsh}), a theory derived from general principles of physics and two assumptions about the basic behavior of  granular media, 
is constructed to qualitatively account for a wide range of observation 
-- from elastic waves over elasto-plastic motion to rapid dense flow. 
In this paper, showing the close resemblance of results from Barodesy and {\sc gsh}, we further validate {\sc gsh} and provide an understanding for  {\sc gr}.   
\end{abstract}

\maketitle

\tableofcontents
\section{Introduction}

One focus of soil mechanical experiments is the stress evolution $\sigma _{ij}\left( t\right) $ for given strain rate $v_{ij}\equiv\frac12(\nabla_iv_j+\nabla_jv_i)$ and density $\rho$~\cite{gudehus2010}. Three striking characteristics being observed at slow, elasto-plastic rates are (1)~rate-independence, (2)~the existence of a critical state~\cite{wood1990}, and (3)~proportional paths as summed up by the Goldscheider Rule ({\sc gr})~\cite{goldscheider}. Rate-independence means that if the given strain rate $v_{ij}$ is a constant, the stress $\sigma_{ij}(t)$ is a function of the strain $\varepsilon_{ij}\equiv\int v_{ij}{\rm d}t=v_{ij}t$, and does not depend on  the actual rate. 

The critical state is an expression of ``ideal plasticity.'' Starting from an isotropic stress $\sigma_{ij}=P\delta_{ij}$, and applying a constant shear rate $v_{ij}^*$ ($^*$ denotes the traceless part) -- while maintaining a vanishing trace $v_{\ell\ell}=0$ to keep the density $\rho$  constant -- a granular system will always go into an asymptotic, stationary state, in which the stress $\sigma^c_{ij}$ no longer changes with time, although $v_{ij}$ goes on providing a constant rate of deformation. We shall call this asymptotic state -- characterized by the direction of the rate $v_{ij}^*/v_s$ (where $v_s^2=v_{ij}^*v_{ij}^*$) and the density $\rho$ -- the {\em critical state}. 
(The asymptotic state is more typically arrived at for given shear rate $v_{ij}^*$, at constant pressure $P\equiv\frac13\sigma_{\ell\ell}$ or one of the principle stress value $\sigma_{11}$, rather than the density. And there are some in the engineering community who insist on restricting the term {\em critical state} to the results of this second type of approaches. The narrower definition would be sensible if the respective asymptotic states were different. We do not believe this to be the case, for rather basic reasons, as will become clear in section~\ref{sec1b}.) 

The Goldscheider Rule  or {\sc gr}  is a generalization of the critical state. First, it states that a granular system will converge onto the same critical state associated with $v_{ij}^*/v_s$ and $\rho$, starting from any initial stress, not only an isotropic one. And second, it postulates the existence of asymptotic states also for cases of changing $\rho$ -- a point that we believe may be understood as follows: 
In the principal strain axes  $(\varepsilon_{11},\varepsilon_{22},\varepsilon_{33})$, a constant $v_{ij}^*$ means the system moves with a constant rate along its direction, $\varepsilon_{11}/\varepsilon_{22}=v_{11}^*/v_{22}^*$, $\varepsilon_{22}/\varepsilon_{33}=v_{22}^*/v_{33}^*$. This circumstance is referred to as {\em a proportional strain path}. In the stress space, $\sigma^c_{ij}$ is a stationary dot and does not move. 
Now, adding  a constant  $v_{\ell\ell}$ to the isochoric strain path, $[v_{ij}^*+\frac13 v_{\ell\ell}\delta_{ij}]\,t$, we need to keep $v_{\ell\ell}$ small compared to $v_s$, such that an initial stress has sufficient time to converge without breaching the random closest or loosest packing -- the grains get crushed in the first case, and loose contacts with one another in the second. Then one expects  the asymptotic state to be approximately given by the critical stress  $\sigma^c_{ij}(\rho)$ associated with the same isochoric strain path $v_{ij}^*\,t$. As time passes,  the density $\rho$ will change, so will the critical state  $\sigma^c_{ij}(\rho)$. But it will remain associated with $v_{ij}^*\,t$. Interestingly,  {\sc gr} states that this stress path is also proportional,  meaning $\sigma_{11}(t)/\sigma_{22}(t)$, $\sigma_{22}(t)/\sigma_{33}(t)$ also remain constant.

The statements of {\sc gr} as rendered by Kolymbas~\cite{barodesy} are: (1)~Proportional strain paths starting from the stress
$\sigma_{ij}=0$ are associated with proportional stress paths.
(2)~Proportional strain paths starting from 
$\sigma_{ij}\not=0$
lead asymptotically to the corresponding proportional
stress paths obtained when starting at 
$\sigma_{ij}=0$. (A caveat: Although Goldscheider is a prudent and reliable experimenter, his data base is rather small~\cite{goldscheider}.)
The initial value $\sigma_{ij}=0$ is a mathematical idealization, neither easily realized nor part of the empirical data that went into {\sc gr}. So we take its core statement as: Applying a proportional strain path, there is an asymptotic line to which all other stresses converge. This stress path is also proportional, such that proportional strain and stress paths form pairs.

Barodesy~\cite{barodesy} is a recent, impressively realistic constitutive model by Kolymbas, the originator of hypoplasticity~\cite{kolymbas1}. His purpose is to further improve the quantitative account of  granular media's motion, and to reduce the considerable liberty when constructing hypoplastic models. He did this by starting explicitly from {\sc gr}. In the present paper, we take Barodesy (as specified below, in Sec~\ref{sec2}), as a constitutive  equation with a reduced set of variables, which reflects highly condensed and intelligently organized empirical data, to which the results of {\sc gsh} are  compared. 

{\sc gsh} is a theory of continuum mechanics derived from two notions that we hold to be the basic physics of granular media~\cite{granR2}.  When constructing the theory employing  general principles, we had little experimental data in mind, and certainly never needed to choose a subset of these. Therefore, if not totally wrong, {\sc gsh} should be adequate and correct in a broad-ranged fashion. Until now, {\sc gsh} has shown itself capable of accounting for phenomena as diverse as static stress distribution~\cite{ge-1,ge-2,granR1}, incremental stress-strain relation~\cite{SoilMech}, yield~\cite{JL2,3inv}, propagation and damping of elastic waves~\cite{ge4},
elasto-plastic motion~\cite{JL3}, the critical state~\cite{critState}, shear
band and fast dense flow~\cite{denseFlow}. Comparison to Barodesy is a further hard test for {\sc gsh},  especially because any agreement could not possibly have been planned for. Moreover, {\sc gsh} provides an understanding for  {\sc gr}, and embeds it among the  many granular phenomena already understood within the framework of  {\sc gsh}.   

Finally, we amplify on the point why comparing  {\sc gsh} successfully to Barodesy validates the former. First, we note the qualitative difference between a physicist’s theory and an engineering
one, which are in fact constructs of different {\em raison d'\^ etre}: Physicists aim to first of all gain a qualitative understanding of a given system, while engineers want primarily to organize and mathematically condense experimental data gained in that system. 
When constructing a theory, physicists typically start from some notions about the basic behavior of a system, calling them motivation. If broadly validated, physicists will conclude these notions are right, considering this a gain in understanding. Two theories with different notions will contradict each other eventually, even if they initially agree with respect to some experiments. Two engineering theories may look very different mathematically, but if essentially the same set of
data was used in constructing these theories, they will not contradict each other starkly -- though small deviations will generally exist. 
An agreement between   {\sc gsh} and Barodesy, showing that a theory deduced from some notions, hence quite possibly wildly off the mark, produces results that are seen in experiments, is therefore indeed a validation. 
On the other hand, there are always shades of gray -- engineers with a fundamentalist's heart and physicists with a most pragmatic mindset. But all will agree that a mature theory should be both realistic and derived from the specific physics of the system under focus.

We shall in future compare   {\sc gsh} to more constitutive models and experiments on elasto-plastic motion. Recent experiments include  uniaxial tests~\cite{aaa10}, the settlement~\cite{aaa8}, and systematic triaxial measurements~\cite{aaa2}, though we do not expect either models or experiments to deviate strongly from Barodesy or each other, for the reasons mentioned above. A recent paper~\cite{aaa1} stands out, because it connects the constitutive model to particle-level properties. Generally speaking, one needs to heed the cautionary words by Schwedes~\cite{aaa11}, that these setups are typically designed for engineering purpose, and the data may depend on operational details (such as skill of the experimenters and their level of training). So care has to be taken when adopting their data. A major difficulty, we believe, comes from the influence of the initial state, from the lack of information on boundary conditions, and from the presence of water (that {\sc gsh} not yet considers). 

DEM simulations are nowadays a popular approach employed by both physicists and engineers, see eg.~\cite{aaa5,aaa6,aaa7,aaa9,aaa12,aaa13,aaa14,aaa15,aaa16,aaa17,aaa18}.
The best are often qualitatively perfect, but numerically different from experimental results. Therefore, a comparison to {\sc gsh} requires us to treat them as different systems, with their own values for energy and transport coefficients. Unfortunately, the associated  calibration process is time-consuming and laborious.

\section{The  Equations of GSH\label{sec1b}}
\subsection{Two-Stage Irreversibility}
The essence of granular physics, we contend, is encapsulated by two notions: {\em two-stage irreversibility} and {\em variable transient elasticity}. The first is related to the three spatial scales of any
granular media: (a)~the macroscopic, (b)~the intergranular, and (c)~the
inner granular. Dividing all degrees of freedom into these three
categories, we treat those of (a) differently from (b,c). Macroscopic
degrees of freedom, such as the slowly varying stress or velocity fields, are
specified and employed as explicit state variables, but intergranular and
inner granular degrees are treated summarily: Instead of being specified,
only their contribution to the energy is considered and taken,
respectively, as granular and true heat. So we do not account for the motion of a jiggling grain, only include its strongly fluctuating kinetic and elastic energy as contributions to the granular 
heat $\int T_g{\rm d}s_g$, characterized by the granular entropy $s_g$ and temperature $T_g$. Similarly, a phonon, or any elastic vibration within the grain, are taken as part of true heat, $\int T{\rm d}s$. Clearly, there are only a handful of macroscopic degrees of freedom (a), innumerable intergranular ones (b), and yet many orders of magnitude more inner granular ones (c). So the statistical tendency to equally distribute the energy among all degrees of freedom implies that the energy decays from (a) to (b,c), and from (b) to (c), never backwards. This is what we call {\em two-stage irreversibility. } 

Accounting for higher densities, when enduring contacts abound and granular jiggling is small, we expand $w_T$ to obtain $w_T=s_g^2/2b$, with $T_g\equiv\partial w_T/\partial s_g=s_g/b$ and $w_T\sim T_g^2$. There is no linear term because $s_g,T_g=0$ is an energy minimum. (The usual granular temperature $T_G$, defined as 2/3 of a grain's average kinetic energy, is useful only in the dilute limit, when the fluctuating elastic energy may be neglected, and $w_T\sim T_G\sim T_g^2$.) Neglecting nonuniform situations, the balance equation for $s_g$  reads
\begin{equation}
T_g{\partial_t s_{g}}=\eta_g v_s^2+\xi_g v_{\ell\ell}^2-\gamma T_{g}^2,
\label{120409-3}
\end{equation}
with $v_s^2\equiv v_{ij}^\ast v_{ij}^\ast$, and $v_{ij}^\ast$ the traceless part of $v_{ij}$. Also: $\partial_t\equiv\frac{\partial}{\partial_t}$. The first two term on the right side accounts for viscous heating, the third for the leak of granular heat from (b) to (c). The viscosities $\eta_g,\xi_g$ and the relaxation rate $\gamma$ are parameters of {\sc gsh} and functions especially of $T_g,\,\rho$.  We take~\cite{granR2}
\begin{equation}
T_g=s_g/(\rho b),\quad \eta_g=\eta_1 T_g,\quad \gamma=\gamma_0+\gamma_1T_g,
\end{equation}
noting that for what we call the {\em hypoplastic regime} of slightly elevated $T_g$, in which hypoplasticity and Barodesy hold, $\gamma_0\ll\gamma_1T_g$ may be neglected. And we neglect $\xi_g v_{\ell\ell}^2$, because $\eta_g v_s^2\gg\xi_g v_{\ell\ell}^2$ in all typical experiments. For constant shear rate $v_{ij}^\ast $, Eq~(\ref{120409-3}) is a relaxation equation, with $T_g$ quickly settling into its stationary value,
\begin{equation}\label{eq3}
T_g=\sqrt{\eta_1/\gamma_1}\, v_s.
\end{equation}
It is then no longer independent. The coefficients $1/b,\eta_1,\gamma_1$ are functions of $\rho$, 
\begin{equation}
b\sim(\rho_{cp}-\rho)^{0.1},\quad \gamma_1,\,\eta_1\sim(\rho_{cp}-\rho)^{-1}, 
\end{equation}
where $\rho_{cp}$ is the closest packed density. We stand behind the $T_g-$dependence with much more confidence than that of the density, for two reasons: First, there are no comparably general arguments to extract the $\rho$ dependence. Second, probably because of this, the observed dependence is not universal. The above dependence fits glass beads data, while 
$ \gamma_1\sim(\rho_{cp}-\rho)^{-0.5}$, $\eta_1\sim(\rho_{cp}-\rho)^{-1.5}$ 
seem more suitable for polystyrene beads, see~\cite{denseFlow}.   

\subsection{Variable Transient Elasticity}

Our second notion, {\em variable transient elasticity}, addresses granular elasticity and plasticity. The free surface of a granular system at rest can be inclined, or tilted. When
perturbed, when the grains jiggle and $T_g\not=0$, the  inclination will be reduced until the surface is horizontal. The stronger the grains jiggle, the faster this process is. We take this as indicative of a system that is elastic for $T_g=0$, turning transiently elastic for $T_g\not=0$, with a stress relaxation rate that grows with $T_g$. A relaxing stress is typical of any viscous-elastic system such as polymers. The
unique circumstance here is that the relaxation rate is not a material
constant, but a function of the state variable $T_g$. As we shall see, it
is this dynamically controlled, {\em variable transient elasticity} -- a simple fact at
heart -- that underlies the complex behavior of granular plasticity.
Realizing it yields a most economic way to capture granular rheology.

Employing a strain field rather than the
stress as a state variable usually yields a simpler description, because the former is in
essence a geometric quantity, while the latter contains material parameters such as stiffness. Yet one cannot use the
standard strain field $\varepsilon_{ij}$ as a granular state variable, because the relation between stress and $\varepsilon_{ij}$ lacks uniqueness when the system is plastic.
A number of engineering theories divide the strain into two fields, elastic and plastic,
$\varepsilon_{ij}=u_{ij}+\varepsilon^{(p)}_{ij}$, with the first accounting for
the reversible and second for the irreversible part. They then employ $\varepsilon_{ij}$ and $\varepsilon^{(p)}_{ij}$ as two independent strain fields to account for granular plasticity~\cite{Houlsby}.

We believe that one should, on the contrary, take the elastic strain $u_{ij}$
as the sole state variable, as there is a unique relation between $u_{ij}$ and the elastic stress $\sigma_{ij}$ -- { if both are related via the elastic energy}: Shearing a granular system, part of the strain goes into deforming the grains, changing their elastic energy. The rest is spent sliding
and rolling the grains. Taking $u_{ij}$ as the portion that changes the
energy and deforms the grains, the elastic energy $w(u_{ij})$ is by definition a function
of $u_{ij}$ alone. And since an elastic stress $\sigma_{ij}(u_{ij})$ only exists when the
grains are deformed, it is also a function of $u_{ij}$. Therefore, we
employ $u_{ij}$ as the sole state variable, and discard both
$\varepsilon_{ij}$ and $\varepsilon^{(p)}_{ij}$. Doing so preserves many useful
features of elasticity, especially the (so-called hyper-elastic) relation,
\begin{equation}\label{1-1} \sigma_{ij}=-\partial w(u_{ij},\rho)/\partial u_{ij}|_\rho.
\end{equation} 
This is derived in~\cite{granR2} but easy to understand via
an analogy. Driving up a snowy hill slowly, the car wheels will grip the ground  part of the time, slipping otherwise. (We assume a slowly turning wheel and quickly changing, intermittent stick-slip behavior.) When the wheels do grip, the car moves upward and its gravitational energy $w^{grav}$ is increased. If we divide the wheel's rotation into a gripping (e) and a slipping (p) portion, $\theta=\theta^{(e)}+\theta^{(p)}$, we know we may ignore $\theta^{(p)}$, and compute the torque on the wheel as $\partial 
w^{grav}/\partial\theta^{(e)}$, same as Eq.(\ref{1-1}). How much the wheel turns or slips, how large $\theta$ or $\theta^{(p)}$ are, is irrelevant for the torque. 

The functional dependence of the energy density $w(u_{ij})$ is an input in {\sc gsh}, as it cannot be obtained from general principles. The one we propose, because we find it both simple and appropriate, is specified below, in Sec~\ref{sec2c}. Once it is given, so is the elastic stress, for which there is therefore an explicit expression, in terms of the state variables $u_{ij}$ and $\rho$. 

The evolution equation for $u_{ij}$, as derived in~\cite{granR2}, may be divided into that of the trace $\Delta\equiv-u_{\ell\ell}$ and the traceless part $u_{ij}^\ast $, 
\begin{eqnarray}
{\partial_t u_{ij}^{\ast }}&=&\left( 1-\alpha \right)
v_{ij}^{\ast }-\lambda T_{g}u_{ij}^{\ast },  \label{120409-1} \\
{\partial_t \Delta }&=&(1-\alpha)v_{\ell\ell} +\alpha
_{1}u_{lk}^{\ast }v_{lk}^{\ast }-\lambda _{1}T_{g}\Delta.  \label{120409-2} 
\end{eqnarray}
Comparing them to the fully elastic equation, ${\partial_t u_{ij}}=v_{ij}$, we realize $\alpha$ is a gear shift factor, as a higher rate is necessary to achieve the same deformation; while $\alpha_1$ is a dilatancy factor, accounting for the granular phenomenon that a shear flow leads to compression or decompression.  Both $\alpha$ and $\alpha_1$ are off-diagonal Onsager coefficients that depend on $T_g$, though we may take them as constant in the present context.  

 The two terms $\sim T_g$ are relaxation terms, accounting for the loss of deformation $\Delta, u_{ij}^{\ast }$ (and the associated loss of the stress) when $T_g$ is finite. The relaxation rate grows with $T_g$, and is typically about 3 times as large for $u_{ij}^{\ast }$ as $\Delta$, hence $\lambda\approx3\lambda_1$~\cite{JL3}. This is how {\em variable transient elasticity} is mathematically encoded.  Replacing $T_g$ with the shear rate $v_s$ for the stationary case of  Eq.(\ref{eq3}), we find the above two equations explicitly rate independent. Denoting $\lambda T_g=\lambda\sqrt{\eta_1/\gamma_1}v_s\equiv\Lambda v_s$, $\lambda_1T_g\equiv\Lambda_1 v_s$, we take, as tentatively as before,  
\begin{equation}
\Lambda,\,\Lambda_1,\,\alpha,\,\alpha_1\sim\rho_{cp}-\rho,
\end{equation}  
assuming that the plastic phenomena of relaxation, softening and dilatancy are no longer operative at $\rho=\rho_{cp}$. For a given shear rate $v_{ij}$, Eqs~(\ref{120409-1},\ref{120409-2}) are relaxation equations. Denoting $u_s^2=u_{ij}^{\ast }u_{ij}^{\ast }$, $v_s^2=v_{ij}^{\ast }v_{ij}^{\ast }$, the system will converge onto the rate independent asymptotic value (denoted by a superscript $^c$) of  
\begin{equation}\label{eq8}
u^c_s=\frac{1-\alpha}{\Lambda},\quad \frac{\Delta^c}{u^c_s}=\frac{\alpha _1}{\Lambda _{1}}+\frac{1-\alpha}{u^c_s\Lambda_1}\frac{v_{\ell\ell}}{v_s},
\end{equation}
and $u_{ij}^{\ast }|^c=v_{ij}^{\ast }(u^c_s/v_s)$ implying $u_{ij}|^c$ and $v_{ij}$ have the same orientation and principal axes. Note also for ${v_{\ell\ell}}\ll{v_s}$, the density dependence of ${\Delta^c}/{u^c_s}\approx{\alpha _1}/{\Lambda _{1}}$ cancels.  

\begin{figure}[t]
\includegraphics[scale=.6]{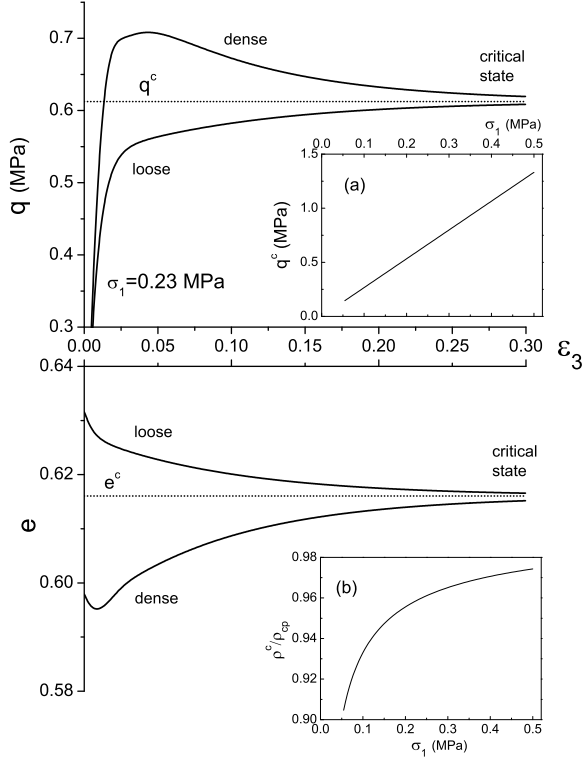}
\caption{Seemingly a textbook illustration of the
critical state, this is the result of a
{\sc gsh}  calculation: Shear stress $q$ and void ratio $e$
against the strain $\varepsilon_3$ in triaxial tests
(cylinder axis along 3), at given $\sigma_1$ and
strain rate $\varepsilon_3/t$, for an initially dense ($\rho_0 =0.97\rho_{cp}$)
and loose ($\rho_0=0.95\rho_{cp}$) sample. Insets are 
critical stress $q^c$ and density $\rho^c$ against
$\sigma_1$; with $B_0 = 7$GPa, ${\cal A} =0.6{\cal B}$,
$\rho_{\ell p}=0.85\rho_{cp}$, $\alpha_1=
750(1-\rho/\rho_{cp})$, $\alpha=0.7$,
$\Lambda=2850(1-\rho/\rho_{cp})$,
$\Lambda/\Lambda_1=3.3$, and in SI units:
$\gamma_1=6\cdot10^5$, $\eta_1=0.15$ (ie.
$\Lambda=114$, $\alpha_1=30$ for $\rho =0.96\rho_{cp}$, as in~\cite{JL3}).
}\label{crit-state}
\end{figure}

Given the elastic strain $u_{ij}|^c=u_{ij}^*|^c-\frac13\Delta^c\delta_{ij}$ and the density $\rho$, the elastic stress is also given. For $v_{\ell\ell}=0$, this 
is simply the ideally plastic, stationary, critical state. The elastic strain and the associated stress do not change with time, because the deformation rate and the relaxation cancel, $\partial_t u_{ij}\sim\partial_t\sigma_{ij}= 0$. Calculating an approach to this state, starting from an isotropic stress and keeping the $\sigma_{1}$ constant, the resultant curves for $q\equiv\sigma_3-\sigma_1=\sigma_s\sqrt{3/2}$ and the void ratio $e\equiv\rho_g/\rho-1$ ($\rho_g$: grain's density), against the strain $\varepsilon_3$ in triaxial tests
(cylinder axis along 3), resemble a textbook illustration of the critical state, see Fig~\ref{crit-state}. 

For $v_{\ell\ell}/v_s$ finite but small, neither the direction nor the magnitude of $u_{ij}$ and $\sigma_{ij}$ change much, as conjectured in the introduction, though the mean stress will grow or decrease with the density. To see how it changes, and why it follows a proportional path, we need the explicit expression for the stress, specified in the next section, Sec~\ref{sec2c}. But we can already see the reason for {\sc gr}'s second statement: Starting from any initial value $u_{ij}\not=u_{ij}|^c$, the deviation  $u_{ij}-u_{ij}|^c$ will, according to Eqs.(\ref{120409-1}) and (\ref{120409-2}),  vanish exponentially. Clearly,  {variable transient elasticity} is what lies behind both the critical state and {\sc gr}. 
 
Finally, we note that the Cauchy stress, or total stress, $\sigma_{ij}^{tot}$ is softer than the elastic one,
\begin{equation}\label{tot}
\sigma_{ij}^{tot}=(1-\alpha)\sigma_{ij},
\end{equation}
with the same factor as in Eqs.(\ref{120409-1}) or (\ref{120409-2}). This is a result of the Onsager reciprocity relation~\cite{LL5}, but it may less formally also be seen as the flip side of the gear shift factor: when a higher rate is needed to achieve a certain elastic  deformation, energy conservation requires the restoring force to be smaller by the same factor.

For very high rates -- such as present in heap flow or avalanches, the seismic pressure $P_T\sim T_g^2\sim v_s^2$ resulting from violent jiggling of the grains, and the viscous stress,  $\eta_g v_{ij}^*=\eta_1T_g v_{ij}^*$, again of shear rate squared, become relevant and destroy the rate independence. We neglect both in this paper -- though not the viscous granular heating of Eq.(\ref{eq3}).

\subsection{Yield versus the Critical State\label{secYvC}}
In a space spanned by stress components and density, there is a surface  that 
divides two regions in any granular media, one in which the grains necessarily move, 
another in which they may be at rest. This surface may be referred to as {\em the 
yield surface}.  Equivalently, we may take the yield surface  as the divide between two regions, one in which elastic solutions are stable, and another in which they are not -- clearly, the medium may be at rest for a given stress only if an appropriate elastic solution is stable. 
Since the elastic energy of any solution satisfying force equilibrium $\nabla_j\sigma_{ij}=0$ is an extremum~\cite{granR2}, the energy is convex and minimal in the stable region, concave and maximal in the unstable one ---in which infinitesimal perturbations suffice to destroy the state of rest.

Yield is clearly a completely distinct concept from the critical state as discussed above -- one a static phenomenon, the other dynamic.  The first is a convexity transition of the elastic energy, to be probed by quasi-static motion at vanishing $T_g$, say by slowly tilting a plate with a layer of grains. The second is a stationary solution of  the relaxation equation for the elastic strain $u_{ij}$ at given shear rate, and relevantly, at an elevated $T_g$. It is comparable to the stationary solution of any diffusion equation. The yield and critical shear stresses are frequently similar in magnitude, but the yield stress  needs to be larger than the highest shear stress achieved during the approach to the critical state. Otherwise, the system will abandon the approach and develop shear bands instead, see Fig.\ref{Stationary-Surface-in-PaiPlane} below. 

Many believe that the approach to the critical state is accomplished at low enough shear rates to be
considered {\em quasi-static}. We contend that a {quasi-static motion} is one that visits a series of static, equilibrium states, with $T_g\to0$. The rate of dissipation must be negligibly small.  The rate-independent,  hypoplastic motion taking place during an approach to the critical state maintains an elevated $T_g$ that allows irreversible, dissipative relaxation of the elastic strain. In the critical state, this dissipative process, having the same magnitude as the reactive (ie. elastic) deformation rate, certainly cannot be neglected.   

\subsection{Granular Elastic Energy\label{sec2c}}
The elastic energy density $w$ is a function of the three independent strain invariants, 
$\Delta, u_{s}$ and $u_{t}\equiv \sqrt[3]{u_{ik}^{\ast
}u_{kj}^{\ast }u_{ji}^{\ast }}$. For granular materials, the following expression is appropriate in many respects,
\begin{equation}
w={\cal B}\sqrt{\Delta }\left( \frac{2}{5}\Delta ^{2}+\frac{1}{\xi }u_{s}^{2}-\frac{%
\chi }{\xi }\frac{u_{t}^{3}}{\Delta }\right),  \label{w}
\end{equation}%
and was instrumental in achieving the agreements with all the granular phenomena mentioned above, especially static stress distribution, incremental stress-strain relation, and elastic waves. Varying the coefficient $\chi$, the yield surface changes to resemble different yield laws, including Drucker-Prager, Lade-Duncan, Coulomb, and Matsuoka-Nakai, see~\cite{3inv}. For qualitative considerations, however, it is frequently sufficient to set $\chi=0$. We then find the elastic energy convex only for 
\begin{equation}\label{2b-3} u_s/\Delta\le\sqrt{2\xi}, \quad
\text{or}\quad \sigma_s/P\le\sqrt{2/\xi},
\end{equation} 
where $P\equiv\sigma _{kk}/3$,
$\sigma _{s} \equiv\sqrt{\sigma _{lk}^{\ast }\sigma _{lk}^{\ast }}$.
The energy $w$ turns concave if this condition is violated. 

We keep a finite $\chi$ for the rest of this paper, taking it along with $\xi\approx5/3$ as density independent. But $\cal B(\rho)$ is specified as~\cite{granR2}
\begin{equation}
\label{2b-4} {\cal B}={\cal B}_0
[(\rho-\bar\rho)/(\rho_{cp}-\rho)]^{0.15}, 
\end{equation}%
with $\bar\rho\equiv(20\rho_{\ell p}-11\rho_{cp})/9$, and ${\cal B}_0>0$. This expression accomplishes three things:
\textbullet~The energy is concave for any density smaller than the
random loose one $\rho_{\ell p}$, implying no elastic solution exists there. 
\textbullet~The energy is convex between the random loose density $\rho_{\ell p}$ and the random close one $\rho_{cp}$, ensuring the stability of any elastic
solutions in this region. In addition, the density dependence of sound
velocities as measured by Harding and Richart~\cite{hardin} is well
rendered by $\sqrt{\cal B}$. 
\textbullet~The elastic energy diverges, slowly, at
$\rho_{cp}$, approximating the observation that the system becomes orders of magnitude stiffer there.

The elastic stress
$\sigma_{ij}=-\partial w/\partial u_{ij}$ may be written as 
\begin{equation}
\sigma_{ij}=\frac{\partial w}{\partial \Delta }\delta _{ij}-\frac{1}{u_{s}}%
\frac{\partial w}{\partial u_{s}}u_{ij}^{\ast }-\frac{1}{u_{t}^{2}}\frac{%
\partial w}{\partial u_{t}}\left( u_{ik}^{\ast }u_{kj}^{\ast }-\frac{%
u_{s}^{2}}{3}\delta _{ij}\right),  \label{GSH-13}
\end{equation}%
and calculated employing Eq.(\ref{w}). Using Eqs.(\ref{GSH-13})
it can also be shown that for any isotropic energy $w$, the stress  and elastic strain tensors have same principal directions. And since the critical elastic strain is colinear with the strain rate, all three have the same principal axes asymptotically. The stress eigenvalues $\sigma_{1,2,3}$ are%
\begin{eqnarray}
\sigma_{i} &=&{\cal B}\sqrt{\Delta }\left[ \Delta +\left( \frac{1}{2\xi }-\frac{%
\chi }{\xi }\right) \frac{u_{s}^{2}}{\Delta }\right.  \label{GSH-15} \\
&&\left. +\frac{\chi }{2\xi }\frac{u_{t}^{3}}{\Delta ^{2}}-\frac{2}{\xi }%
\left( u_{i}+\frac{\Delta }{3}\right) +\frac{3\chi }{\xi \Delta }\left(
u_{i}+\frac{\Delta }{3}\right) ^{2}\right],  \notag
\end{eqnarray}%
where $u_{i}$ denote eigenvalues of $u_{ij}$. From (\ref{GSH-15}), the following relations between the triplet of strain invariants $\left( \Delta, u_{s}, u_{t}\right) $ and stress invariants $\left( P,\sigma
_{s},\sigma_{t}\equiv\sqrt[3]{\sigma _{ik}^{\ast }\sigma _{kj}^{\ast }\sigma
_{ji}^{\ast }}\right)$ holds:
\begin{eqnarray}
\frac P{\cal B} &=&{\Delta }^{3/2}\left[ 1+\frac{{u_{s}^{2}}}{2\xi \Delta ^{2}}\left(
1+\chi \frac{{u_{s}}}{{\Delta }}{\frac{u_{t}^{3}}{{u_{s}^{3}}}}\right) %
\right],  \label{a21} \\
\frac {\sigma_{s}}{\cal B} &=&\frac{{\Delta }^{3/2}}{\xi}\frac{u_{s}}{\Delta}\sqrt{4+3{\chi \frac{u_{s}%
}{\Delta }}\left( \frac{{\chi }}{2}{\frac{u_{s}}{\Delta \,}}-4\frac{{%
u_{t}^{3}}}{{u_{s}^{3}}}\right) },  \label{a22} \\
\frac{\sigma_{t}}{\cal B} &=&\frac{{\Delta }^{3/2}}{\xi}\frac{u_{s}}{\Delta}\sqrt[3]{6{\chi }\,\frac{{%
u_{s}}}{{\Delta }}-\left( 8+9{\chi }^{2}\frac{{u_{s}^{3}}}{{\Delta }^{3}}%
\right) {\frac{u_{t}^{3}}{{u_{s}^{3}}}}+3{\chi }^{3}\frac{{u_{s}^{3}}}{{%
\Delta }^{3}}\left( 3\frac{{u_{t}^{6}}}{{u_{s}^{6}}}-\frac{1}{4}\right) }.
\label{a23}
\end{eqnarray}
We are now in a position to understand that the proportional stress path is in fact a result of certain coefficients (or combinations of them) not depending on the density: The above three formulas show that $P/\sigma_s$ and $\sigma_t/\sigma_s$ depend on $\chi,\,\xi,\,\Delta^c/u^c_s,\, u^c_t/u^c_s$. If all four are independent of the density, the stress path is proportional, with the stress magnitude growing with $\rho$, as
\begin{equation}\label{gsh-dichte}
\text{stress\,\, magnitude} \sim(1-\alpha){\cal B}(\rho)\,\, {\Delta }^{3/2},
\end{equation}
cf. Eqs~(\ref{tot},\ref{2b-4}). These four quantities are indeed density independent for $v_1,v_2\gg v_{\ell\ell}$: First, because of $u^c_t/u^c_s=v_t/v_s$, see the first equation after Eq~(\ref{eq8}), or alternatively Eq~(\ref{120418-21}) below, the quantity $u^c_t/u^c_s$ depends only on the direction of the shear rate. Second, $\chi,\,\xi,\,\Delta^c/u^c_s\approx\alpha_1/\Lambda_1$ have been taken above as density independent.

\section{The Barodesy Model\label{sec2}}

In soil mechanics, granular dynamics is frequently modeled employing  the strategy of
{\em rational mechanics}, by postulating an algebraic function $\mathfrak{C}_{ij}$ -- of the  stress $\sigma _{ij}$, strain rate $v_{k\ell}$, and density $\rho $ -- such that the constitutive relation, $(\partial _{t}+v_k\nabla_k)\sigma _{ij}=\mathfrak{C}_{ij}( \sigma_{ij},v_{k\ell},\rho)$ holds.  (Instead of the density, one can equivalently take the void ratio $e\equiv\rho _{bulk}/\rho -1$, with $\rho _{bulk}$ the bulk density of the grains.) Together with the continuity  equation $\partial _{t}\rho +\nabla _{i}\rho v_{i}=0$, momentum conservation, $\partial _{t}(\rho v_{i})+\nabla_{j}(\sigma _{ij}+\rho v_iv_j)=0$, it forms a closed set of equations for $\rho $, $\sigma _{lk}$ and the velocity $v_{i}$. Barodesy as proposed by Kolymbas~\cite{barodesy} is such a model. Note that, in comparison to {\sc gsh}, the state variables $T_g$ and $u_{ij}$ have been eliminated by taking certain limits, such as stationarity or rate-independence. It is therefore a constitutive equation with a reduced number of variables. Barodesy is defined by the following expressions: 
\begin{eqnarray}\nonumber
\partial _{t}\sigma _{ij}&=&c_{0}^{1-c_{3}}\left( \sigma _{mn}\sigma
_{mn}\right) ^{c_{3}/2}\sqrt{v_{lk}v_{lk}}\times
\\
&&\left[ \left( \frac{c_{4}v_{kk}}{%
\sqrt{v_{lk}v_{lk}}}-c_{5}e_{c}\right) \frac{r_{ij}}{\sqrt{r_{mn}r_{mn}}}+%
\frac{c_{5}e\sigma _{ij}}{\sqrt{\sigma _{mn}\sigma _{mn}}}\right],
\label{Barodesy-1}
\end{eqnarray}
where the direction and magnitude of the asymptotic stress are given respectively as
\begin{eqnarray}
r_{ij} &\equiv&\frac{v_{kk}}{\sqrt{v_{lk}v_{lk}}}\delta _{ij}+c_{1}\exp \left(
c_{2}\frac{v_{ij}}{\sqrt{v_{lk}v_{lk}}}\right),  \label{Barodesy-2} \\
\frac{1+e_{c}}{1+e_{0c}} &\equiv&\exp \left[ \left( \frac{\sigma _{mn}\sigma
_{mn}}{c_{0}^{2}}\right) ^{\left( 1-c_{3}\right) /2}\frac{1}{c_{4}\left(
1-c_{3}\right) }\right].  \label{Barodesy-3}
\end{eqnarray}
Similar to {\sc gsh},  the asymptotic stress $r_{ij}$ has the same principal axes as the strain rate $v_{ij}$, ie.  $r_{ij}$ is diagonal if $v_{ij}$ is. There are 6 dimensionless parameters, the values of which are:
\begin{eqnarray}\label{barodesy-parameters}
c_{1} &=&-1.7637,\,\, c_{2}=-1.0249,\,\, c_{3}=0.5517, \label{Barodesy-parameters}
\\
c_{4} &=&-1.174,\,\, c_{5}=-3.26,\,\,e_{0c}=0.75,  \notag
\end{eqnarray}
in addition to one with dimension, $c_0$, in Pa.

\section{Strain Paths of Constant Density}
In this section, we evaluate {\sc gsh} and Barodesy analytically. This is possible only for $v_{\ell\ell}=0$ and  constant density. So we are dealing in fact with the critical state here, as discussed above a special case of  {\sc gr}.  A proportional deformation path is given
by all three eigenvalues of $v_{ij}\left( t\right) $ remaining 
proportional. We decompose the strain rate tensor as 
\begin{equation}
v_{ij}=v\left(
\begin{array}{ccc}
\widetilde{v}_{1} & 0 & 0 \\
0 & \widetilde{v}_{2} & 0 \\
0 & 0 & \widetilde{v}_{3}%
\end{array}%
\right),  \label{120415-1}
\end{equation}%
with $\widetilde{v}_{1}+\widetilde{%
v}_{2}+\widetilde{v}_{3}=0$, $\sqrt{\widetilde{v}_{1}^{2}+\widetilde{v}_{2}^{2}+\widetilde{v}_{3}^{2}%
}=1$, and $v\equiv \sqrt{v_{lk}v_{lk}}>0$ denoting the magnitude. 
Proportional paths imply the constancy of $\widetilde{v}_{1,2,3}$. 
Next, we employ the strain Lode angle,
\begin{equation}
L_{\varepsilon}\equiv\left( 1/3\right) \arcsin \left( \sqrt{6}v_{t}^{3}/v_{s}^{3}\right),
\label{120418-5}
\end{equation}%
(where $v_{t}\equiv \sqrt[3]{v_{ik}^{\ast }v_{kj}^{\ast }v_{ji}^{\ast }}$, and $v_{s}=v$ because $v_{kk}=0$) to express $\widetilde{v}_{1,2,3}$,
\begin{eqnarray}
\widetilde{v}_{1} &=&\sqrt{\frac{2}{3}}\sin \left( L_{\varepsilon} -\frac{\pi }{3}%
\right) ,  \label{120418-2} \\
\widetilde{v}_{2} &=&-\sqrt{\frac{2}{3}}\sin L_{\varepsilon},  \label{120418-3} \\
\widetilde{v}_{3} &=&\sqrt{\frac{2}{3}}\sin \left(L_{\varepsilon} +\frac{\pi }{3}%
\right) .  \label{120418-4}
\end{eqnarray}
Similarly, we can write the stress tensor as%
\begin{equation}
\sigma _{ij}=\sigma \widehat{R}\left(
\begin{array}{ccc}
\widetilde{\sigma }_{1} & 0 & 0 \\
0 & \widetilde{\sigma }_{2} & 0 \\
0 & 0 & \widetilde{\sigma }_{3}%
\end{array}%
\right) \widehat{R}^{T}  \label{120418-7}
\end{equation}%
with $\widehat{R}$ a rotation matrix -- equal to the unit matrix for the asymptotic states in  both {\sc gsh} and barodesy. $\widetilde{\sigma }_{1,2,3}$ need to be expressed by two angles, because $\widetilde{\sigma }_{1}+\widetilde{\sigma }_{2}+\widetilde{\sigma }_{3}\not=0$. We take them as the stress Lode angle $L$ and the friction angle $\zeta $, as defined in the Appendix,  Eqs.(\ref{120417-1},\ref{120417-2}),
\begin{eqnarray}
\widetilde{\sigma }_{1} &=&\frac{\cos \zeta }{\sqrt{3}}+\frac{2\sin \zeta }{%
\sqrt{6}}\sin \left( L-\frac{\pi }{3}\right) ,  \label{120418-8} \\
\widetilde{\sigma }_{2} &=&\frac{\cos \zeta }{\sqrt{3}}-\frac{2\sin \zeta }{%
\sqrt{6}}\sin L,  \label{120418-9} \\
\widetilde{\sigma }_{3} &=&\frac{\cos \zeta }{\sqrt{3}}+\frac{2\sin \zeta }{%
\sqrt{6}}\sin \left( L+\frac{\pi }{3}\right) .  \label{120418-10}
\end{eqnarray}%
Proportional path implies time-independent $L,\zeta$. [The relation between the angles $L, \zeta $ and the stress invariants $P, \sigma _{s}, \sigma _{t}$ are given in the Appendix, Eqs.(\ref{120417-1},\ref{120417-2}).]
The association between the strain and stress paths may be given as
\begin{equation}
L =L\left( L_{\varepsilon} \right)  \label{120418-11} \quad
\zeta =\zeta \left( L_{\varepsilon} \right) 
\end{equation}

We now calculate the stress evolution for proportional
strain paths, 
\begin{equation}
v_{ij}=\sqrt{\frac{2}{3}}v\left(
\begin{array}{ccc}
\sin \left(L_{\varepsilon}  -\frac{\pi }{3}\right) & 0 & 0 \\
0 & -\sin L_{\varepsilon} & 0 \\
0 & 0 & \sin \left( L_{\varepsilon}  +\frac{\pi }{3}\right)%
\end{array}%
\right) \text{,}  \label{120418-13}
\end{equation}
obtained by inserting (\ref{120418-2},\ref{120418-3},\ref{120418-4})
into (\ref{120415-1}), taking $v=$  const. Both   {\sc gsh} and Barodesy deliver analytical expressions. 

\subsection{Results from GSH}

The   {\sc gsh} equations can be solved as follows. First, inserting the strain
rate (\ref{120418-13}) into Eq.(\ref{120409-3}), and noting that $%
v=v_{s}=const$, we have that the solution for $T_{g}$ is:
\begin{eqnarray}
T_{g} &=&v\sqrt{{\eta_1 }/{\gamma _{1}}}\tanh \left( vt\,\sqrt{\gamma
_{1}\eta_1 }\right)  \label{120418-14} \\
&=&v\sqrt{{\eta_1 }/{\gamma _{1}}}\tanh \left(\varepsilon \, \sqrt{\gamma _{1}\eta_1 }%
\right)  \notag
\end{eqnarray}%
Here the initial condition $T_{g}(t=0)=0$ is assumed. The notation $%
\varepsilon =\int_{0}^{t}vdt^{\prime }=vt$ is magnitude of total strain.
Clearly $1/\left( v\sqrt{\gamma _{1}\eta_1 }\right) $ (or $1/\sqrt{\gamma
_{1}\eta_1 }$) is the time (or strain) scale needed for $T_{g}$ going to its
saturation (asymptotic) value.

\begin{figure}[tbh]
\begin{center}
\includegraphics[scale=0.5]{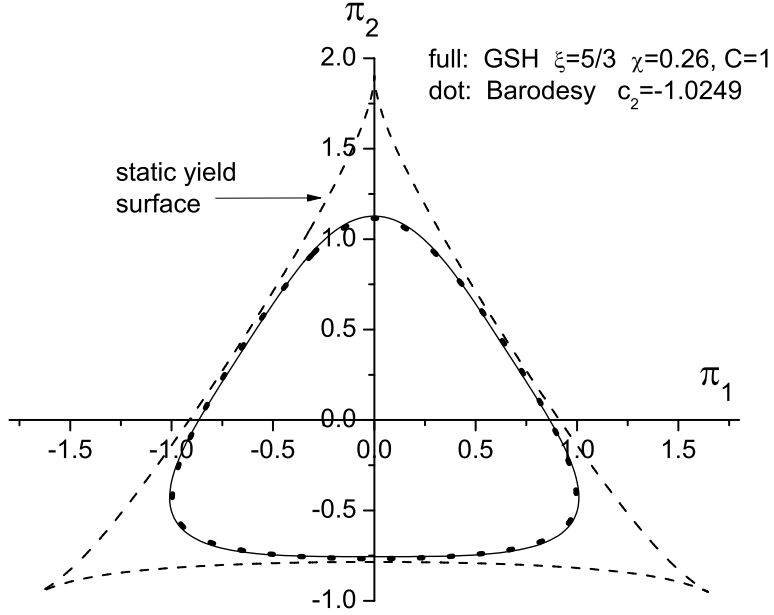}
\end{center}
\caption{Loci of asymptotic  states in the $\protect\pi $-plane, as obtained, respectively, from  {\sc gsh} (full) and
barodesy (dots). Dashed curve is the static yield surface, ie. the convexity surface of the energy $w$, Eq.(\ref{w}).}
\label{Stationary-Surface-in-PaiPlane}
\end{figure}

Inserting (\ref{120418-14}) into Eq.(\ref{120409-1}), we
obtain the deviatoric elastic strain,
\begin{equation}
u_{ij}^{\ast }=\frac{u_{ij0}^{\ast }+\left( 1-\alpha _{s}\right)
v_{ij}^{\ast }\int_{0}^{t}\cosh ^{\lambda /\gamma _{1}}\left( v\sqrt{\gamma
_{1}\eta _{1}}t^{\prime }\right) dt^{\prime }}{\cosh ^{-\lambda /\gamma
_{1}}\left( \sqrt{\gamma _{1}\eta _{1}}\varepsilon \right) }
\label{120418-15}
\end{equation}%
where $u_{ij0}^{\ast }=u_{ij}^{\ast }\left( t=0\right) $ is the initial strain. Because $\cosh x\rightarrow e^{x}$ for $x\rightarrow
\infty $, the initial strain $u_{ij0}^{\ast }$ decays as $%
e^{-\Lambda\varepsilon }$. The dimensionless parameter%
\begin{equation}
\Lambda\equiv\lambda \sqrt{\eta_1/\gamma _{1}}  \label{120418-18}
\end{equation}
is typically $\approx 114$ for the intermediate void ratio of $e=$ 0.65, see~\cite{granR2}. So the decay is fast, ending at about $1\%$  strain
magnitude $\varepsilon $. Inserting (\ref{120418-14}) into (\ref{120409-2}), we have
\begin{eqnarray}
\Delta &=&\frac{\Delta _{0}+\int_{0}^{t}h\left( t^{\prime }\right) \cosh
^{\lambda _{1}/\gamma _{1}}\left( v\sqrt{\gamma _{1}\eta_1 }t^{\prime }\right)
dt^{\prime }}{\cosh ^{-\lambda _{1}/\gamma _{1}}\left( \sqrt{\gamma _{1}\eta_1
}\varepsilon \right) },\qquad \text{ with}  \label{120418-16} \\
h &\equiv &\alpha _{1}\frac{u_{lk0}^{\ast }v_{lk}^{\ast }+\left( 1-\alpha
\right) v^{2}\int_{0}^{t}\cosh ^{\lambda /\gamma _{1}}\left( v\sqrt{\gamma
_{1}\eta_1 }t^{\prime }\right) dt^{\prime }}{\cosh ^{-\lambda /\gamma
_{1}}\left( \sqrt{\gamma _{1}\eta _1}\varepsilon \right) }.  \label{120418-17}
\end{eqnarray}%
So the initial bulk strain $\Delta _{0}$ decays with $e^{-\Lambda
_{1} \varepsilon }$, more slowly, as $\Lambda _{1}\approx\Lambda/3 $.
For $t\rightarrow \infty $, the strain (\ref{120418-15},\ref{120418-16}) becomes
stationary:
\begin{eqnarray}
u_{ij}^{\ast } &\rightarrow &u_{ij}^{\ast }|^c=\frac{1-\alpha }{\Lambda}\frac{%
v_{ij}^{\ast }}{v},  \label{120409-7} \\
\Delta &\rightarrow &\Delta_{c}=\frac{\alpha _{1}\left( 1-\alpha \right)
}{\Lambda \Lambda_1}.  \label{120409-8}
\end{eqnarray}%
The associated three invariants are
\begin{eqnarray}
u^{c}_s =\frac{1-\alpha }{\Lambda},\quad 
\frac{u^{c}_s}{\Delta^c } =\frac{\Lambda_{1}}{\alpha _{1}},
  \label{120418-19}
\\
\frac{u_{t}^{c}}{u^{c}_s} =\frac{v_{t}}{v}=\frac{\sin ^{1/3}\left(
3L_{\varepsilon}\right)}{6^{1/6}},  \label{120418-20}
\label{120418-21}
\end{eqnarray}%
where the third expression is obtained with Eq.(\ref{120418-5}).
Inserting these into (\ref{GSH-15}), we have the following principal stresses:
\begin{eqnarray}
\sigma _{1}^{c} &=&\frac{B}{\xi }\left( \frac{1-\alpha }{\Lambda}C\right) ^{3/2}%
\left[ \xi +\left( \frac{1}{2}-\chi \right) C^{2}\right.  \notag \\
&&+\left. \frac{\chi C^{3}}{2\sqrt{6}}\sin \left( 3L_{\varepsilon}  \right) -2C\sin
\left(L_{\varepsilon}  -\frac{\pi }{3}\right) +3\chi C^{2}\sin \left(L_{\varepsilon}  -\frac{%
\pi }{3}\right) \right]
\end{eqnarray}%
\begin{eqnarray}
\sigma _{2}^{c} &=&\frac{B}{\xi }\left( \frac{1-\alpha }{\Lambda}C\right) ^{3/2}%
\left[ \xi +\left( \frac{1}{2}-\chi \right) C^{2}\right.  \notag \\
&&+\left. \frac{\chi C^{3}}{2\sqrt{6}}\sin \left( 3L_{\varepsilon}  \right) +2C\sin
L_{\varepsilon}  -3\chi C^{2}\sin L_{\varepsilon} \right]
\\
\sigma _{3}^{c} &=&\frac{B}{\xi }\left( \frac{1-\alpha }{\Lambda}C\right) ^{3/2}%
\left[ \xi +\left( \frac{1}{2}-\chi \right) C^{2}\right.  \notag \\
&&+\left. \frac{\chi C^{3}}{2\sqrt{6}}\sin \left( 3L_{\varepsilon} \right) -2C\sin
\left(L_{\varepsilon} +\frac{\pi }{3}\right) +3\chi C^{2}\sin \left(L_{\varepsilon}  +\frac{%
\pi }{3}\right) \right]
\end{eqnarray}%
where $C\equiv \Lambda _{1}/\alpha _{1}$. In the $\pi $-coordinates (defines as $\pi _{1}=({\sigma _{s}}/{P})\sin \left( L+\frac{\pi }{6}\right)$, 
$\pi _{2} =-({\sigma _{s}}/{P})\cos \left( L+\frac{\pi }{6}\right)$, see the Appendix for more details), the critical stress has more compact expressions:
\begin{eqnarray}
\pi _{1} &=&\frac{6C^{2}\chi \sin \left( 2L_{\varepsilon}  +\frac{\pi }{3}\right) -4%
\sqrt{6}C\cos \left(L_{\varepsilon}  -\frac{\pi }{3}\right) }{2\sqrt{6}\xi +\sqrt{6}%
C^{2}+\chi C^{3}\sin 3L_{\varepsilon} },  \label{GSH-22} \\
\pi _{2} &=&\frac{6C^{2}\chi \cos \left( 2L_{\varepsilon}  +\frac{\pi }{3}\right) -4%
\sqrt{6}C\sin \left(L_{\varepsilon}  -\frac{\pi }{3}\right) }{2\sqrt{6}\xi +\sqrt{6}%
C^{2}+\chi C^{3}\sin 3L_{\varepsilon} }.  \label{GSH-23}
\end{eqnarray}%
When the Lode angle $L_{\varepsilon} $ varies from $0$ to $2\pi $, the loci given by
Eqs.(\ref{GSH-22},\ref{GSH-23}) give a triangle-like curve, as shown by the full
line in Fig.\ref{Stationary-Surface-in-PaiPlane}. The curve is
determined by the three parameters: $C$, $\xi $, $\chi $, and reduces to a
circle if $\chi =0$.

\subsection{Results from Barodesy}

The critical surface of the Barodesy is obtained by taking $\partial
_{t}\sigma _{ij}=0$ and $v_{kk}=0$. In this
case,  Eq.(\ref{Barodesy-1}) reduces to  $e=e_{c}$ and 
\begin{equation}
\sigma _{ij}=\frac{\sqrt{\sigma _{mn}\sigma _{mn}}}{\sqrt{\exp \left( \frac{%
c_{2}v_{mn}}{\sqrt{v_{lk}v_{lk}}}\right) \cdot \exp \left( \frac{c_{2}v_{mn}%
}{\sqrt{v_{lk}v_{lk}}}\right) }}\exp \left( \frac{c_{2}v_{ij}}{\sqrt{%
v_{lk}v_{lk}}}\right) .  \label{Barodesy-4}
\end{equation}%
Inserting the strain rate Eq.(\ref{120418-13}) in the principal axes,
 we have
\begin{eqnarray}
\sigma _{1} &=&\left( \sigma _{1}\sigma _{2}\sigma _{3}\right) ^{1/3}\exp
\left[ c_{2}\sqrt{\frac{2}{3}}\sin \left(L_{\varepsilon}  -\frac{\pi }{3}\right) %
\right],  \label{Barodesy-10} \\
\sigma _{2} &=&\left( \sigma _{1}\sigma _{2}\sigma _{3}\right) ^{1/3}\exp
\left( -c_{2}\sqrt{\frac{2}{3}}\sin L_{\varepsilon} \right),  \label{Barodesy-11} \\
\sigma _{3} &=&\left( \sigma _{1}\sigma _{2}\sigma _{3}\right) ^{1/3}\exp
\left[ c_{2}\sqrt{\frac{2}{3}}\sin \left( L_{\varepsilon}  +\frac{\pi }{3}\right) %
\right].  \label{Barodesy-12}
\end{eqnarray}%
or in the $\pi $-coordinates:%
\begin{eqnarray}
\pi _{1} &=&\frac{3}{\sqrt{2}}\frac{1-\exp \left[ \sqrt{2}\left\vert
c_{2}\right\vert \cos \left( L_{\varepsilon}  -\frac{\pi }{3}\right) \right] }{\exp
\left( \sqrt{2}\left\vert c_{2}\right\vert \cos L_{\varepsilon}  \right) +\exp \left[
\sqrt{2}\left\vert c_{2}\right\vert \cos \left(L_{\varepsilon}  -\frac{\pi }{3}%
\right) \right] +1},  \label{Barodesy-20} \\
\pi _{2} &=&\frac{3}{\sqrt{6}}\frac{2\exp \left( \sqrt{2}\left\vert
c_{2}\right\vert \cos L_{\varepsilon} \right) -\exp \left[ \sqrt{2}\left\vert
c_{2}\right\vert \cos \left( L_{\varepsilon}  -\frac{\pi }{3}\right) \right] -1}{\exp
\left( \sqrt{2}\left\vert c_{2}\right\vert \cos L_{\varepsilon}  \right) +\exp \left[
\sqrt{2}\left\vert c_{2}\right\vert \cos \left( L_{\varepsilon}  -\frac{\pi }{3}%
\right) \right] +1}.  \label{Barodesy-21}
\end{eqnarray}%

\begin{figure}[t]
\begin{center}
\includegraphics[scale=0.5]{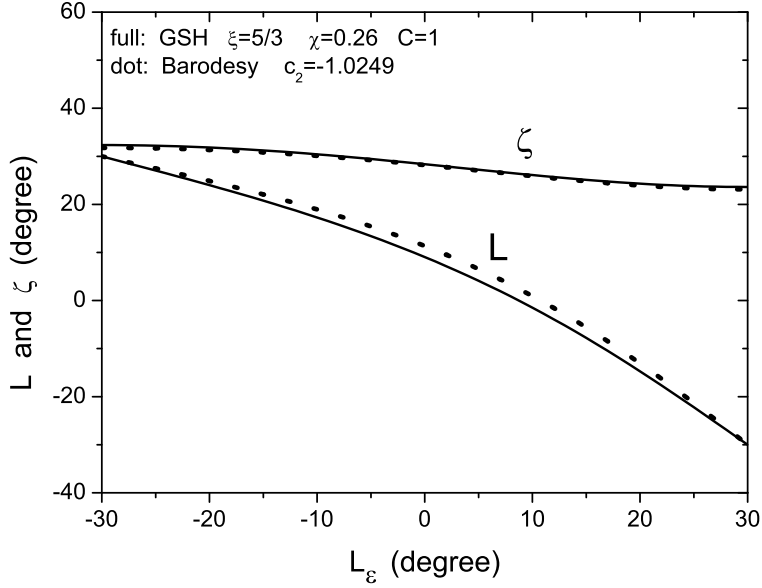}
\end{center}
\caption{ Angle association of Eq.(33), calculated employing   {\sc gsh} (full) Barodesy (dots).}
\label{fig2}
\end{figure}

In barodesy, the critical surface  is determined by the parameter $c_{2}$, and is triangle-like in the $\pi $-plane, see Fig.\ref{Stationary-Surface-in-PaiPlane}, which is the same curve as in Fig.1 of the second reference of~\cite{barodesy}.

Transforming both the {\sc gsh} expressions of Eqs.(\ref{GSH-22},\ref{GSH-23}) and the Barodesy ones of Eq.(\ref{Barodesy-20},\ref{Barodesy-21}) into the angles $L,\zeta $ using the
formula given in the appendix, we retrieve the association of Eq.(\ref{120418-11}), as shown in Fig.\ref{fig2}, again with great similarity between both theories. 

\begin{figure}[tbh]
\begin{center}
\includegraphics[scale=0.7]{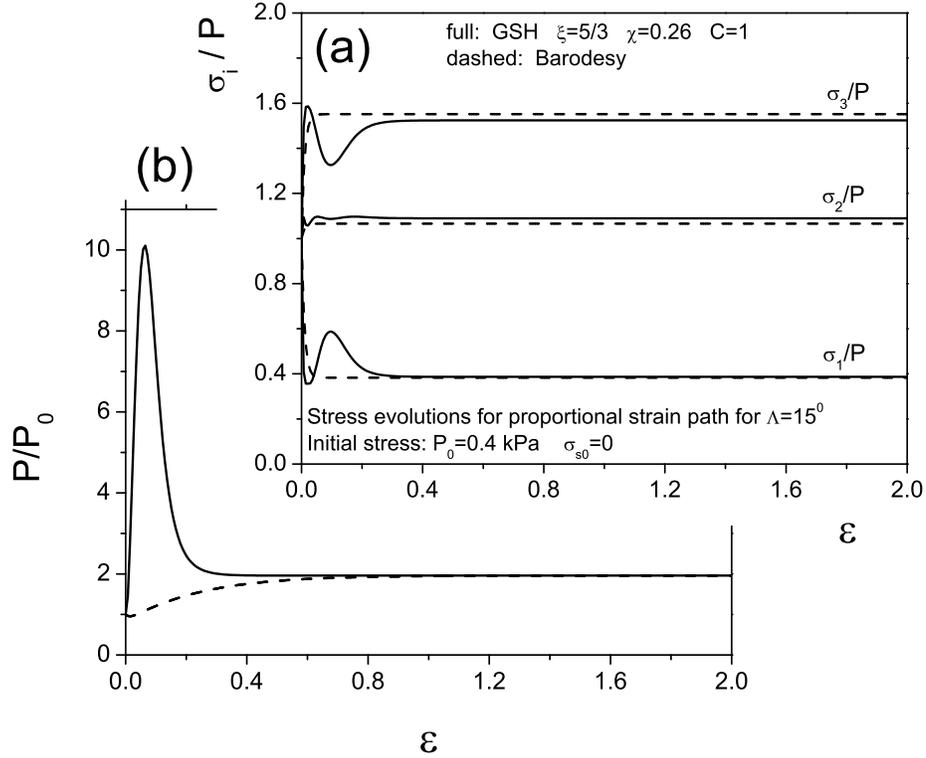}
\end{center}
\caption{(a)~Evolution of the stress eigenvalues (normalized by the pressure $P$) with the total strain $\varepsilon$, as obtained with
 {\sc gsh} (full) and Barodesy (dashed). (b)~Evolution of $P$ (normalized by the initial pressure $P_0$) with $\varepsilon$. The Barodesy curve is monotonic, The {\sc gsh} one is not. This explains the difference in the transient regime.}
\label{fig3}
\end{figure}

In contrast to the last two figures that contain only asymptotic  information, Fig.\ref{fig3} shows the evolution of three stress eigenvalues,  starting from an initially isotropic stress state. The numeric calculation employs {\sc gsh}, Eqs.(\ref{120409-1},\ref{120409-2},\ref{120409-3}) and Barodesy, Eqs.(\ref{Barodesy-1}), for $L_{\varepsilon}  =15^{0}$. The transient behavior is clearly somewhat different, it contains an oscillation in  {\sc gsh} (full), but is monotonic in  Barodesy (dashed). The discrepancy is probably  due to the (correct) nonmonotonic behavior of the pressure and shear stress in {\sc gsh}. 

Although the expressions from barodesy, Eqs.(\ref{Barodesy-20},\ref{Barodesy-21}), and {\sc gsh}, Eq.(\ref{GSH-22},\ref{GSH-23}), are rather different, the relevant plots are not. Yet to achieve this agreement, hardly any fiddling with the parameters was
necessary. The Barodesy parameters were simply taken from Eq.(\ref{barodesy-parameters}); the {\sc gsh} parameters are essentially the same as we employed them before: $\xi, \chi$ are part of the energy and represent static parameters. We took $\xi=5/3$ as we have mostly done before, and took $\chi=0.26$. (In~\cite{3inv}, we took $\chi=0.2$. This slight change perfected the agreement of Fig.1 that we could not resist.) We also took  
$C\equiv \Lambda\lambda _{1}/\left(\lambda \alpha _{1}\right) =1$ here. Previously, we equivalently took $\Lambda\equiv \lambda \sqrt{\eta _{1}/\gamma _{1}}\sim 114$, $\lambda
_{1}\sim 3\lambda /10$, and $\alpha _{1}\sim 33$ in~\cite{JL3,critState}, separately.

\section{Strain Paths with density Change}
\begin{figure}[tbh]
\begin{center}
\includegraphics[scale=0.5]{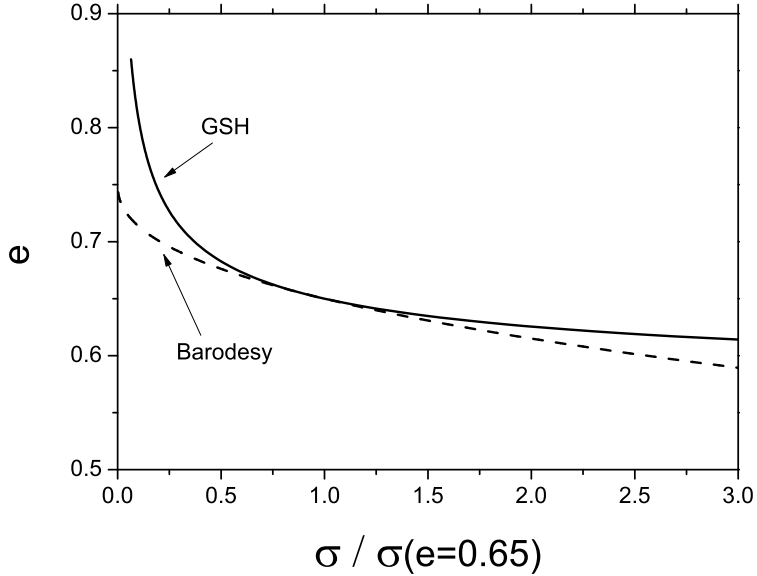}
\end{center}
\caption{The relation between the void ratio $e$ and the stress magnitude $\sigma$  of the asymptotic state  (normalized by  $\sigma$ at $e=0.65$), as given respectively by {\sc gsh} and barodesy. With $v_{\ell\ell}\not=0$, the two quantities $e$ and  $\sigma$ will change in tandem, according to either of these two curves.}
\label{fig6}
\end{figure}

\begin{figure}[tbh]
\begin{center}
\includegraphics[scale=0.59]{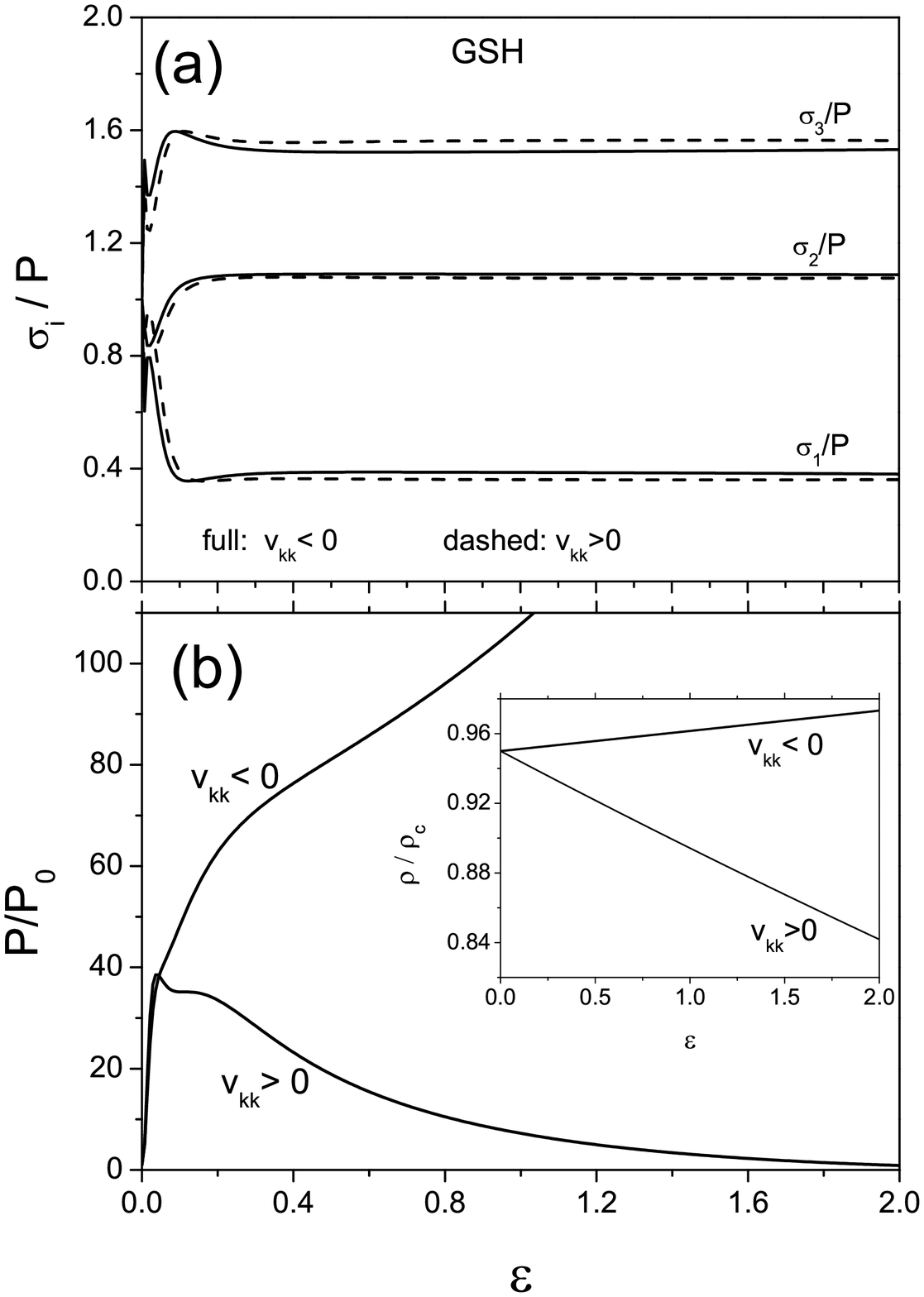}
\end{center}
\caption{Circumstances are the same as in Fig.3, except that,  first,  $v_{kk}\not=0$ and the density varies linearly, as given in the inset of (b). Second, only the {\sc gsh} curve is displayed, not the Barodesy one (because given Fig.\ref{fig3},\ref{fig6}, they cannot be that different). Convergence of the stress ratios takes place quickly, from where on the stress path is proportional. The first part of the pressure change occurs during the convergence, the second part, where the pressure change associated with positive and negative $v_{kk}$ diverge, belongs to the asymptotic state. }
\label{fig7}
\end{figure}

\begin{figure}[tbh]
\begin{center}
\includegraphics[scale=0.85]{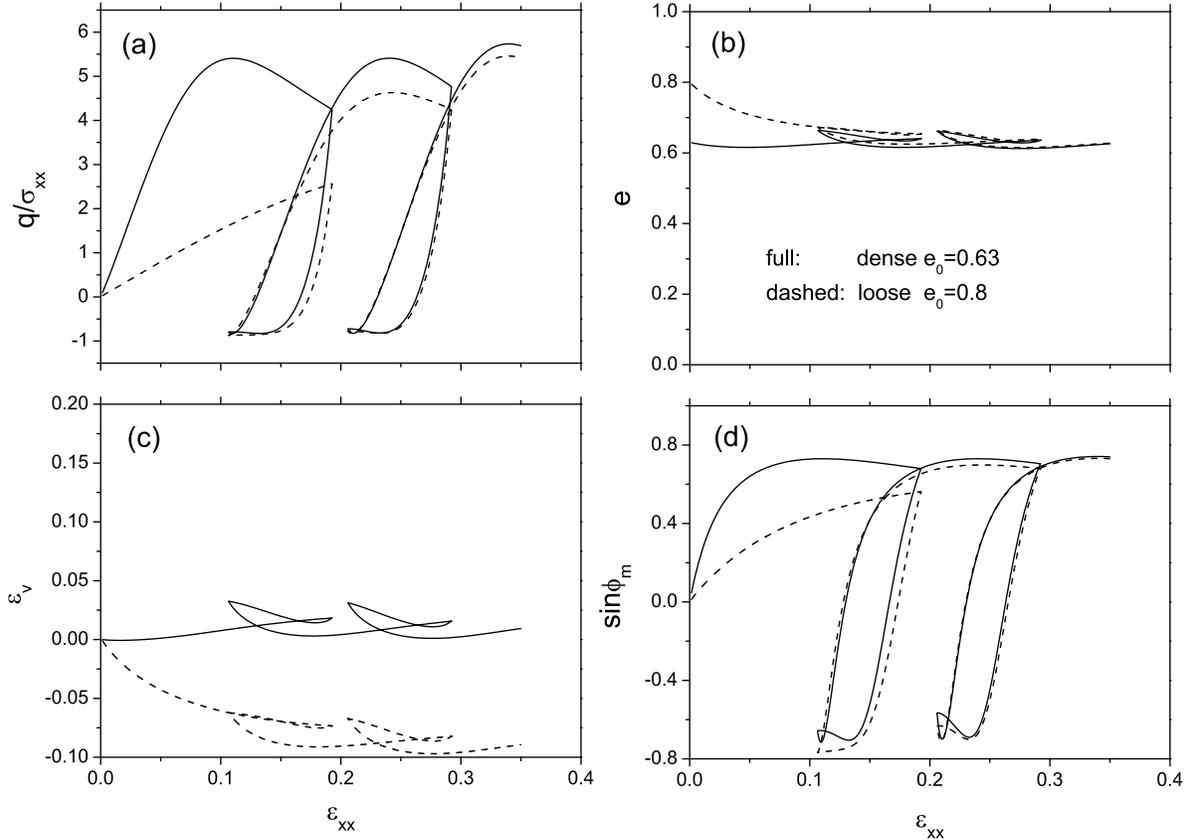}
\end{center}
\caption{In the geometry of triaxial tests, various quantities are  
computed employing {\sc gsh}, as functions of the strain $\varepsilon_{xx}$, holding $\sigma_{xx}=\sigma_{yy}$ constant.  (The axial direction is $z$.  The case with an initially higher density is rendered in solid lines, the looser one in dashed lines.) These are: (a)~deviatoric stress $q\equiv\sigma_{zz}-\sigma_{xx}$; (b)~void ratio $e$; (c)~volumetric strain $\varepsilon_v$; (d) the friction ange, $\sin \phi _{m}\equiv q/\left( 2\sigma_{xx}+q\right) $. We chose: $\alpha ,\alpha _{1},\lambda \sim \left( 1-\rho /\rho
_{cp}\right) ^{1.6}$ and $\eta_1,\gamma_1 \sim \left( 1-\rho /\rho _{cp}\right) 
^{-1}$.}
\label{fig9}
\end{figure}

\begin{figure}[tbh]
\begin{center}
\includegraphics[scale=0.85]{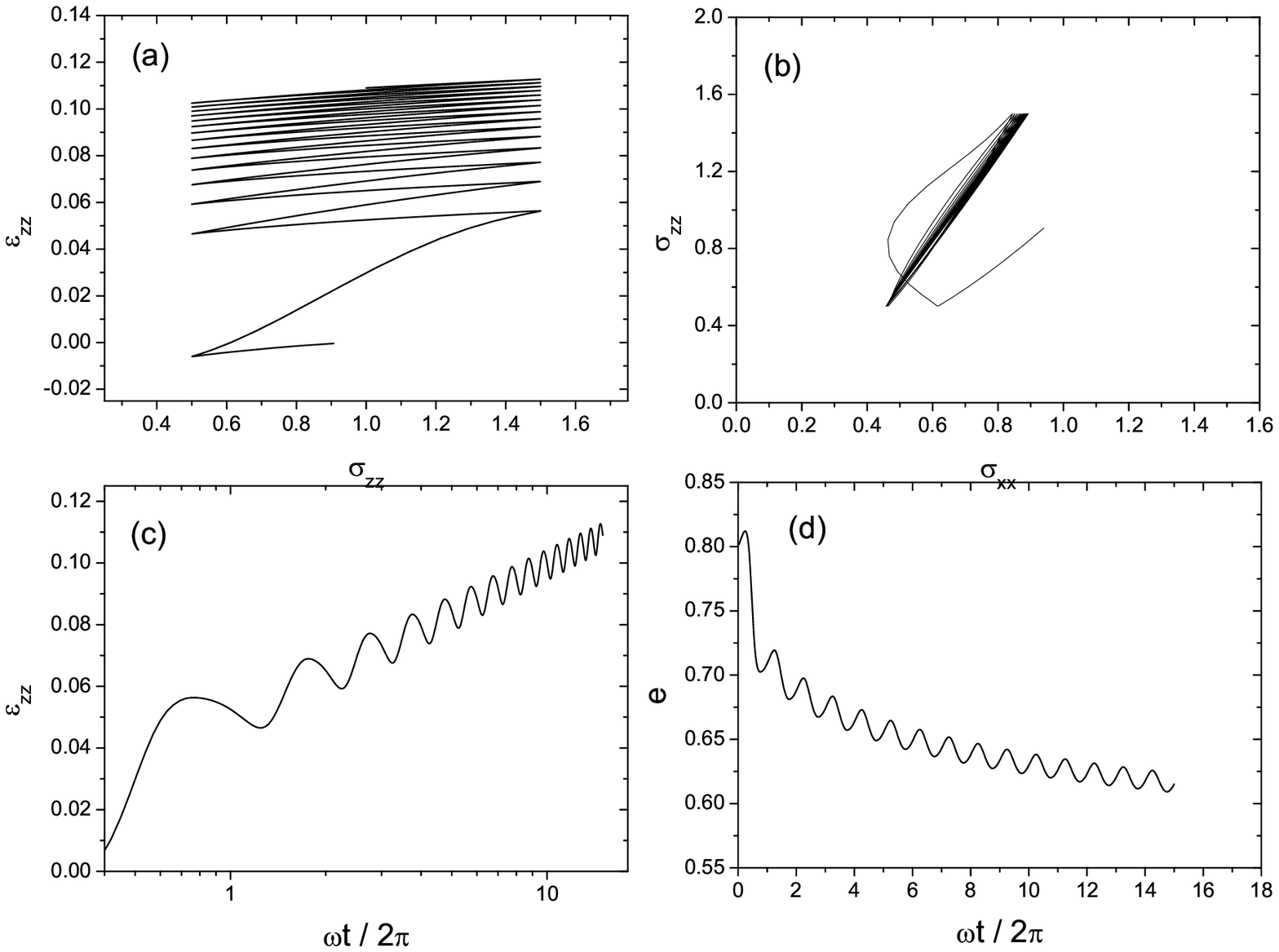}
\end{center}
\caption{Oedometric test starting with  an initially loose density, again computed with {\sc gsh}. The following four quantities are calculated as functions of the axial $\sigma_{zz}$ while holding the lateral strain $\varepsilon_{xx}=\varepsilon_{yy}$ fixed: (a) $\varepsilon_{zz}$ of a axial stress-strain curve; (b) $\sigma_{xx}$ of a stress path; (c)  $\varepsilon_{zz}$ versus time; (d) void ratio $e$ versus time. Note $\varepsilon_{zz}$ moving upwards implies a compaction of the system.}
\label{fig8}
\end{figure}

If the strain path contains a small $v_{\ell\ell}$, the density and void ratio will change, as will the magnitude of the stress, according 
to Eq.(\ref{gsh-dichte}) in {\sc gsh}, and to Eq.(\ref{Barodesy-3}) in barodesy.  Again, in spite of the different expressions, the curves are similar, at least qualitatively, see Fig.(\ref{fig6}). The convergence onto the asymptotic state is depicted in Fig.(\ref{fig7}). Following Kolymbas' papers, we have also computed 4 figures each for (drained) triaxial and oedometric tests employing {\sc gsh}, see Fig.\ref{fig9} and \ref{fig8}. 

\section{Conclusion}

In comparing {\sc gsh} to barodesy, we set out to achieve two goals: To validate {\sc gsh}, and to provide a transparent, sound understanding for {\sc gr}.  Both goals were reached. {\sc gsh} is validated because it yields similar results for various key quantities as barodesy,  achieving  better agreement than would be reasonably expected, without much fiddling with parameters.   

Conversely, the understanding of Goldscheider Rules and Barodesy comes from the physics of {\sc gsh}. The theory has (for the range of shear rates typical of soil mechanical experiments) three state variables and  two constituent parts. The state variables are the density $\rho$, granular temperature $T_g$ (quantifying granular jiggling), and the elastic strain tensor $\hat u$ (accounting for the coarse-grained deformation of the grains). The two constituent parts are first the explicit expression of the stress tensor $\hat \sigma$, a function of $\hat u, \rho$ that is obtained from the elastic energy; and second a rate-independent relaxation equation for $\hat u$, derived from the notion of {variable transient elasticity}. Given any initial  $\hat u_0$, the system will always converge onto the stationary solution $\hat u^c$ as prescribed by the relaxation equation.   $\hat u^c$ is a function of the constant strain rate, or equivalently, of the proportional strain path's direction, and may be identified as the asymptotic, critical state for isochoric paths, $v_{\ell\ell}=0$. This convergence, a consequence of the relaxation equation, is closely related to { variable transient elasticity},  and hence a generic aspect of granular behavior. 

Given  $\hat u^c$ and the density, the stress is also fixed. Its form, however, depends on the expression for the elastic energy that is material dependent and less robust. 
If the strain path ${\hat v^*}\,t$ is isochoric, with $v_{\ell\ell}=0$, 
the asymptotic stress state is a constant of time, but a function of $\hat u^c$, or equivalently, of the strain path's  direction. As the path varies, the associated stress states lie within a triangle, as depicted in Fig.\ref{Stationary-Surface-in-PaiPlane}. 

If the shear rate is a sum of  ${\hat v^*}\,t$ and a small  $v_{\ell\ell}$, the asymptotic state is (cum grano salis) still given by the $\hat u^c$ associated with ${\hat v^*}\,t$, though the density will now change. The asymptotic stress is therefore a function of the same $\hat u^c$ and a changing density, hence no longer a constant. That the stress path is also proportional, that only the magnitude of the stress changes with time, not the ratios of its eigenvalues, is the least robust part of {\sc gr}, because it depends on the density dependence of certain coefficients canceling.  

Constructing a constitutive relation, specifying $(\partial_t+v_k\nabla_k)\hat \sigma={\mathfrak C}(\hat\sigma,\hat v,\rho)$, is only for someone with vast experience with granular media and deep knowledge of how they behave. That {\sc gsh} -- derived from two simple notions of what the two basic elements of granular physics are -- yields an equivalent account, is eye-opening, and the actually amazing fact of the presented agreement.

\acknowledgements {Helpful discussion with, and critical reading of the manuscript by, Dimitris Kolymbas are gratefully acknowledged.}

\appendix
\section{Tensor decomposition\label{TensorDecop}}

A $3\times 3$ symmetric tensor, e.g. the stress tensor $\sigma _{ij}$, can
be decomposed into two parts: a spatial rotation $\widehat{O}$ and a part
which is invariant under any rotation. In most analysis we are interested
mainly in the three invariants. There are usually various ways to
represent the invariant triplet. One of which is 
\begin{eqnarray}
P &\equiv &\sigma _{kk}/3,  \label{120417-0} \\
\sigma _{s} &\equiv &\sqrt{\sigma _{lk}^{\ast }\sigma _{lk}^{\ast }},
\label{120417-0a} \\
\sigma _{t} &\equiv &\sqrt[3]{\sigma _{ik}^{\ast }\sigma _{kj}^{\ast }\sigma
_{ji}^{\ast }}.  \label{120417-0b}
\end{eqnarray}
Another is $\left( P,L,\zeta \right) $ where
\begin{eqnarray}
L &\equiv &\frac{1}{3}\arcsin \left( \sqrt{6}\frac{\sigma _{t}^{3}}{\sigma
_{s}^{3}}\right) ,  \label{120417-1} \\
\zeta &=&\arctan \left( \frac{\sigma _{s}}{\sqrt{3}P}\right) =\arcsin \left(
\frac{\sigma _{s}}{\sigma }\right) =\arccos \left( \sqrt{3}\frac{P}{\sigma }%
\right),  \label{120417-2}
\end{eqnarray}%
are two angle variables ($\sigma \equiv \sqrt{\sigma _{lk}\sigma _{lk}}=%
\sqrt{\sigma _{s}^{2}+3P^{2}}$). In soil mechanics $L$ is usually called the Lode angle of stress. The angle $\zeta $ can be interpreted as a "friction
angle" (because it represents the ratio between shear force and pressure).
Moreover we can also use the three eigenvalues $\left( \sigma _{1},\sigma
_{2},\sigma _{3}\right) $ of the stress tensor as an invariant triplet,
which are related to $\left( P,\sigma _{s},\sigma _{t}\right) $ by%
\begin{eqnarray}
\sigma _{1} &=&P+\frac{2\sigma _{s}}{\sqrt{6}}\sin \left( L-\frac{\pi }{3}%
\right) ,  \label{120417-3} \\
\sigma _{2} &=&P-\frac{2\sigma _{s}}{\sqrt{6}}\sin L,  \label{120417-4} \\
\sigma _{3} &=&P+\frac{2\sigma _{s}}{\sqrt{6}}\sin \left( L+\frac{\pi }{3}%
\right) ,  \label{120417-5}
\end{eqnarray}%
where $L$ is given by (\ref{120417-1}). In soil mechanics, it is also usual 
to define the two coordinates $\left( \pi _{1},\pi _{2}\right) $ in the so called $\pi
$-plane, 
\begin{eqnarray}
\pi _{1} &=&\frac{\sigma _{3}-\sigma _{2}}{\sqrt{2}P},  \label{120417-6a} \\
\pi _{2} &=&\frac{2\sigma _{1}-\sigma _{2}-\sigma _{3}}{\sqrt{6}P}.
\label{120417-6b}
\end{eqnarray}%
Inserting (\ref{120417-3},\ref{120417-4},\ref{120417-4}) into (\ref%
{120417-6a},\ref{120417-6b}), we have
\begin{eqnarray}
\pi _{1} &=&\frac{\sigma _{s}}{P}\sin \left( L+\frac{\pi }{6}\right),
\label{120417-6} \\
\pi _{2} &=&-\frac{\sigma _{s}}{P}\cos \left( L+\frac{\pi }{6}\right).
\label{120417-7}
\end{eqnarray}%
With the help of Eqs.(\ref{120417-0}-\ref{120417-7}) we can readily
transform among the invariant triplets: $\left( P,\sigma _{s},\sigma
_{t}\right) $, $\left( P,L,\zeta \right) $, $\left( \sigma _{1},\sigma
_{2},\sigma _{3}\right) $, $\left( P,\pi _{1},\pi _{2}\right) $.
Similar decompositions apply for the elastic strain $u_{ij}$, total strain ,
strain rate tensor $v_{ij}$ etc., only note that the first invariant is
frequently defined with a factor different from that of $P$.

\end{document}